\documentstyle[12pt]{article}

\begin{document}

\title{Minimal $SO(10)$ Unification\footnote{Supported in part by 
Department of Energy Grant Number DE FG02 91 ER 40626 A007, and 
DOE/ER 01545-714}}

\author{{\bf S.M. Barr}
\\Bartol Research Institute\\
University of Delaware\\Newark, DE 19716\\ \\
{\bf Stuart Raby}\\
Department of Physics\\Ohio State University\\
Columbus, OH 43210}

\date{BA-97-18\\OHSTPY-HEP-T-97-008}
\maketitle

\begin{abstract}

It is shown that the doublet-triplet splitting problem can be solved
in $SO(10)$ using the Dimopoulos-Wilczek mechanism with a very
economical Higgs content and simple structure. Only one adjoint
Higgs field is required, together with spinor and vector fields.
The successful SUSY GUT prediction of gauge coupling unification 
is preserved. Higgsino-mediated proton decay can be
suppressed below (but not far below) present limits without fine-tuning. 

\end{abstract}

\section{Introduction}

The striking unification of gauge couplings$^1$ at about
$10^{16}$ GeV in the minimal supersymmetric standard model (MSSM)
points toward the possibility of a supersymmetric grand unified 
theory (SUSY GUT). 
The minimal SUSY $SU(5)$ prediction of $\sin^2 \theta_W$ is
$0.2334 \pm 0.0036$, while the experimental value is $0.2324 \pm 0.0003$.$^1$

$SO(10)$ is generally thought to be the most 
attractive grand unified group for a number of reasons. It achieves
complete quark-lepton unification for each family, explains the 
existence of right-handed neutrinos and of ``see-saw" neutrino masses,
has certain advantages for baryogenesis, in particular since
$B-L$ is broken$^2$, and has the greatest promise for explaining the 
pattern of quark and lepton masses.$^3$ 

The greatest theoretical problem that any grand unified theory
must face is the gauge hierarchy problem,$^4$ and in particular that
aspect of it called the ``doublet-triplet splitting problem"$^5$, or
2/3 splitting problem for short. The only way to achieve natural
2/3 splitting in $SO(10)$ is the Dimopoulos-Wilczek (DW) mechanism.$^6$
In Ref. 7 it was shown that realistic $SO(10)$ models can be 
constructed using the DW mechanism.

One criticism that is sometimes made about $SO(10)$ is that solving
the 2/3 splitting problem requires a somewhat
complicated Higgs structure. The models constructed in Ref. 7 and 8 
contained at least the following Higgs 
multiplets: three adjoints (${\bf 45}$), two rank-two symmetric
tensors (${\bf 54}$), a pair of spinors (${\bf 16} + \overline{{\bf 
16}}$), two vectors (${\bf 10}$), and several singlets. Aside from the 
issue of simplicity, there are some indications that 
it is difficult to construct grand unified models with a multiplicity 
of adjoint fields from superstring theory.$^9$

The necessity of a complicated Higgs structure is largely 
traceable to one technical problem, 
namely the breaking of the rank of $SO(10)$ from five to four
without destabilizing the DW solution of the gauge hierarchy
problem. The complete breaking of $SO(10)$ to the 
Standard Model requires at least two sectors of Higgs: 
an adjoint sector and a spinor sector. The adjoint sector plays
the central role in the DW mechanism for 2/3 splitting, while
the spinor sector both breaks the rank of $SO(10)$ and gives
right-handed neutrinos mass.  

The dilemma is that if the spinor sector is coupled to the
adjoint sector it tends to destabilize the DW form of the 
adjoint vacuum expectation value required for the 2/3 splitting,
while if the two sectors are {\it not} coupled (or coupled very 
weakly$^{10,11}$) to each other in the superpotential there arise colored 
and charged (pseudo)goldstone fields due to the fact that that certain
generators of $SO(10)$ are broken by both the adjoint and the spinors,
and in particular the generators that transform as $(3,2,\frac{1}{6})$,
$(\overline{3}, 2, - \frac{1}{6})$, $(\overline{3}, 1, - \frac{2}{3})$,
and $(3,1, \frac{2}{3})$. These destroy the unification of couplings.$^{10}$

In Refs. 7 and 8, an indirect way to couple the two sectors together
without destabilizing the hierarchy was found. However, this 
solution to the problem involved a somewhat complicated Higgs
structure including at least three adjoint fields. 

In this letter we show that there is a very simple way to couple 
the spinor and adjoint sectors together, with a stable hierarchy and 
no pseudo-goldstones. Before describing it, it will be helpful to review 
in more detail the problems that have been discussed above.

\section{The problem of a stable hierarchy in $SO(10)$}

The DW mechanism is based on the existence of an adjoint Higgs field, 
which we shall call $A$, getting
a vacuum expectation value (VEV) in the $B-L$ direction:

\begin{equation}
\langle A \rangle = \left( \begin{array}{ccccc}
0 & & & & \\ & 0 & & & \\ & & a & & \\ & & & a & \\ & & & & a
\end{array} \right) \otimes i \tau_2,
\end{equation}

\noindent
where $a \sim M_G$, the unification scale.
The lower-right $3 \times 3$ block corresponds to $SU(3)$
of color, and the upper-left $2 \times 2$ block to Weak $SU(2)$. 
When this adjoint is coupled to vector 
representations, which we will call $T_1$ and $T_2$, by terms
such as $T_1 \; A \; T_2$ the color-triplets in the vectors
are given superlarge masses, while the Weak doublets remain massless.
(By having also a term of the form $M_T T_2^2$ it is ensured that
only one pair of Weak doublets remains light. The proton-decay
amplitude from the exchange of colored Higgsinos is proportional
to $M_T/a^2$, so that if $M_T/a \stackrel{_<}{_\sim} 10^{-1}$ it
is sufficiently suppressed.$^{7,8}$) Such a ``DW form" for the 
adjoint VEV is not possible in $SU(5)$, where tr$(A) = 0$. 

An adjoint alone is not sufficient to break $SO(10)$ to the 
Standard Model group, $G_{SM}$, and in particular cannot
reduce the rank of the group. This requires
either spinorial Higgs (${\bf 16} + \overline{{\bf 16}}$) or 
rank-five antisymmetric tensor Higgs (${\bf 126} + \overline{{\bf 126}}$). 
As the latter tend to destroy the perturbativity of the unified
interactions below the Planck scale, we will assume that the 
rank-breaking sector has spinors, which we shall call $C$ and 
$\overline{C}$. These spinors (also necessary to give mass to the 
right-handed neutrinos) have VEVs in the $SU(5)$-singlet direction.

The spinor VEVs break $SO(10)$ down to $SU(5)$, and thus the sector
of the superpotential which depends on $C$ and $\overline{C}$ but
not on $A$, which we shall call $W_C$, leaves massless at least those
components of the spinors in the coset $SO(10)/SU(5)$. That is
just a ${\bf 10} + \overline{{\bf 10}} + {\bf 1}$ of $SU(5)$, or a 
$[ (3, 2, \frac{1}{6}) + (\overline{3}, 1, - \frac{2}{3}) +
(1,1, +1) + {\rm H.c.}] + (1,1,0)$ of $G_{SM}$.

The adjoint $A$, with
the VEV shown in Eq.(1), breaks $SO(10)$ down to $SU(3)_c \times
U(1)_{B-L} \times SO(4)$. ($SO(4) = SU(2)_L \times SU(2)_R$.) 
The part of the superpotential that depends on $A$ but not on the 
spinors, which we shall call $W_A$, thus leaves massless at least 
those components of the adjoint in the cosets $SO(10)/(SO(6) \times 
SO(4))$ and $SO(6)/(SU(3)_c \times U(1)_{B-L})$. The first of these 
cosets consists of a $({\bf 6}, {\bf 4})$ of $SO(6) \times SO(4)$, 
which contains $[ (3, 2, \frac{1}{6}) + (3,2 -\frac{5}{6}) + 
{\rm H.c.}]$ of $G_{SM}$. The second coset consists of $[(\overline{3}, 
1, - \frac{2}{3}) + {\rm H.c.}]$ of $G_{SM}$. 

Thus, with no coupling between the two sectors, there are extra, 
uneaten goldstone fields in $(3, 2, \frac{1}{6}) + (\overline{3}, 
1, - \frac{2}{3}) + {\rm H.c.}$. To avoid these, the adjoint must
couple to the spinor. The obvious way to couple them together,
by the term $g \overline{C} \; A \; C$, directly destabilizes 
the DW form assumed for $\langle A \rangle$. Let us assume that 
$A$ has the form ${\rm diag}(b,b,a,a,a) \otimes i \tau_2$. $b = 0$
is the desired DW form. The VEVs of the spinors point in the 
$SU(5)$-singlet direction and have magnitude $c_0 \sim M_{GUT}$. 
Then $g \overline{C} \; A \; C = - \frac{g}{2}
(2b + 3a) c_0^2$. The terms $W_A$ must have a form that gives
$\partial W_A/\partial b = O(M_{GUT}) b$, so that by themselves they
would give $b = 0$. But taking into account also the coupling term
$g  \overline{C} \; A \; C$, one has $0 = -F_b^* = \partial W_{tot}/
\partial b = O(M_{GUT}) b - g c_0^2$, or $b \sim g M_{GUT}$. 
In the DW mechanism$^7$ the Higgs doublets get a ``see-saw" mass of 
order $b^2/M_{GUT}$, so that $g$ must be less than about $10^{-7}$. 
This is the assumption made in Ref. 11. This leads to the 
pseudo-goldstone fields getting masses only of order $g M_{GUT}
< 10^9$ GeV, and thus to $\sin^2 \theta_W = 0.2415$.

The spinor sector and adjoint sector must be coupled together
in some more subtle way. In Ref. 7 such a way was proposed. 
There the spinor sector was assumed to contain a different adjoint
Higgs, called $A'$, whose VEV points in the SU(5)
singlet direction. The two sectors (namely the $A$ sector and the
$(C, \overline{C}, A')$ sector) were then coupled together by
a term ${\rm tr} A A' A^{\prime \prime}$, where $A^{\prime \prime}$
was a third adjoint. Because the $A A' A^{\prime \prime}$ is
totally antisymmetric under the interchange of any two adjoints
(due to the fact that the adjoint is an antisymmetric tensor), there 
have to be three distinct adjoints in this term. This antisymmetry
ensures, as it is easy to see, that this term does not contribute
to any of the $F$ terms as long as the VEVs of the three adjoints commute
with each other. Therefore it does not destabilize the DW form of
the VEV of $A$. And yet, it can also be shown that this trilinear
term is sufficient to prevent the existence of any pseudo-goldstone 
fields.

This has been the choice until now: to assume a complicated Higgs
sector with at least three adjoint Higgs fields$^{7,12}$ or to assume 
that light pseudo-goldstones exist which disturb the beautiful 
SUSY GUT prediction of the Weak angle.$^{10,11,13}$

\section{Solving the problem}

The solution to the above difficulty turns out to be remarkably
simple. Let there be a single adjoint field, $A$, and {\it two} pairs 
of spinors, $C + \overline{C}$ and $C' + \overline{C}'$. The 
complete Higgs superpotential is assumed to have the form

\begin{equation}
W = W_A + W_C + W_{ACC'} + (T_1 A T_2 + S T_2^2).
\end{equation}

\noindent
The precise forms of $W_A$ and $W_C$ do not matter, as long as $W_A$ 
gives $\langle A \rangle$ the DW form, and $W_C$ makes the VEVs of 
$C$ and $\overline{C}$ point in the $SU(5)$-singlet direction. For 
specificity we will take $W_A = \frac{1}{4 M} {\rm tr} A^4 + \frac{1}{2} 
P_A ({\rm tr} A^2 + M_A^2) + f(P_A)$, where $P_A$ is a singlet, $f$ is an 
arbitrary polynomial, and $M \sim M_G$. (It would be possible, also, to have 
simply $m {\rm tr} A^2$, instead of the two terms containing $P_A$.
However, explicit mass terms for adjoint fields may be difficult to
obtain in string theory.$^9$) We take $W_C =
X(\overline{C} C - P_C^2)$, where $X$ and $P_C$ are singlets, and
$\langle P_C \rangle \sim M_G$.

The crucial term that couples the spinor and adjoint sectors
together has the form

\begin{equation}
W_{ACC'} = \overline{C}' \left( \left( \frac{P}{M_P} \right) A + Z 
\right) C + \overline{C} \left( \left( \frac{\overline{P}}{M_P} \right)
A + \overline{Z} \right) C',
\end{equation}

\noindent
where $Z$, $\overline{Z}$, $P$, and $\overline{P}$ are singlets. 
$\langle P \rangle$ and $\langle \overline{P} \rangle$ are assumed 
to be of order $M_G$. The critical point is that the VEVs of the 
primed spinor fields will vanish, and therefore the terms in Eq. (3)
will not make a destabilizing contribution to $- F_A^* = 
\partial W/\partial A$. This is the essence of the mechanism.

$W$ contains several singlets ($P_C$, $P$, $\overline{P}$, and $S$)
that are supposed to acquire VEVs of order $M_G$, but which are
left undetermined at tree-level by the terms so far written down. 
These VEVs may arise radiatively when SUSY breaks, or may be fixed
at tree level by additional terms in $W$, possible forms for which
will be discussed below. 

The VEVs of $A$ and $P_A$ are determined by the equations 
$0 = - F_A^* = \frac{1}{M} A^3 + P_A A$ and $0 = - F_{P_A}^* =
\frac{1}{2}({\rm tr} A^2 + M_A^2) + f'(P_A)$. If $\langle A \rangle =
{\rm diag} (a_1, a_2, a_3, a_4, a_5) \otimes i \tau_2$, then the
first equation implies that $a_i^2 = 0$ or $M \langle P_A
\rangle \equiv a^2$, for each $i$. There is, therefore, a discrete
vacuum degeneracy. The DW vacuum is obtained if two of the $a_i$'s 
vanish and the other three have the same sign and magnitude $a$.
In that case, ${\rm tr} A^2 = - 6 M \langle P_A \rangle$ and 
$\langle P_A \rangle$ is determined by $0 = f'(\langle P_A \rangle )
- 3 M \langle P_A \rangle + M_A^2/2$.

$F_X = 0$ implies that $\langle \overline{C} C \rangle = \langle P_C
\rangle^2 \sim M_G^2$. The $D$ terms and soft, SUSY-breaking terms
will ensure that $\langle \overline{C} \rangle \cong \langle C \rangle
\sim M_G$.

The most interesting equations are 

\begin{equation}
0 = - F_{\overline{C'}}^* = \left( \left( \frac{P}{M_P} \right) A + 
Z \right) C,
\end{equation}

\noindent
and

\begin{equation}
0 = -F_{C'}^* = \overline{C} \left( \left( \frac{\overline{P}}{M_P}
\right) A + \overline{Z} \right).
\end{equation}

\noindent
It is only necessary to consider the first 
of these two equations, as they have the same structure. Let the VEV 
of $C$ be decomposed as follows: $\langle C \rangle = \sum_K f_K C_K$, 
where the $C_K$ are the irredicible multiplets of $G_{SM}$ and the $f_K$ 
are numerical coefficients. Since we have chosen the DW form for 
$\langle A \rangle$, Eq.(4) can be written

\begin{equation}
\left( \frac{3}{2} a \left( \frac{P}{M_P} \right) (B-L)_K + Z \right) 
f_K = 0, 
\end{equation}

\noindent
for all $K$. Since $\langle \overline{C} C \rangle \neq 0$, not all 
the $f_K$ vanish. Suppose $f_J$ does not vanish. Then $Z$ is fixed
to be $Z = - \frac{3}{2} a \left( \frac{P}{M_P} \right) (B-L)_J$. This, 
in turn, implies that $f_K = 0$ for all $K$ for which
$(B-L)_K \neq (B-L)_J$. There are, therefore, a discrete number
of solutions; in fact, four. One of them is $Z = - \frac{3}{2} a
\left( \frac{P}{M_P} \right)$, with $\langle C \rangle$ pointing in 
any direction which has $B-L = 1$. There is a two-complex-dimensional 
space of such directions spanned by the ``$e^+$" and ``$N^c$" directions; 
that is, in terms of $G_{SM}$ quantum numbers, the $(1,1, +1)$ and 
$(1,1,0)$ directions. But actually these are all gauge-equivalent.
This is easily checked directly, but is also clear from the fact,
which shall presently be seen, that there are no uneaten goldstone
modes in this model. Thus, we can take the VEV of $C$ to lie in the
$SU(5)$-singlet direction without any loss of generality.

Now we will show that $\langle C' \rangle = \langle \overline{C}'
\rangle = 0$.
That $C'$ and $\overline{C}'$ have no VEV in the $SU(5)$-singlet
direction follows from $F_Z = F_{\overline{Z}} = 0$. From the
$SU(5)$-singlet component of the $F_C$ and $F_{\overline{C}}$ 
equations one has that $X=0$. And from the $SU(5)$-non-singlet
components of the same equations it follows that $C'$ and 
$\overline{C}'$ have no VEVs in the $SU(5)$-non-singlet directions,
either. All VEVs have now been fixed except for those of $P_C$,
$P$, $\overline{P}$, and $S$, about which more will be said below.

Knowing the VEVs, one can now read off the Higgsino mass matrices 
directly from $W$. For the representations $K = (3,2, \frac{1}{6})$, 
$(\overline{3}, 1, - \frac{1}{3})$, and $(1,1, +1)$, which are
contained in the ${\bf 10}$ of $SU(5)$, one has
$3 \times 3$ mass matrices, since such representations exist in
the adjoint $A$ and in the spinors $C$ and $C'$. The masses come from 
the terms in Eq.(3), both through the VEVs of $A$, $Z$ and $\overline{Z}$,
and through the VEVs
of the spinors $C$ and $\overline{C}$.

\begin{equation}
W_{{\rm mass, 10}}(K) = 
\left( A_{\overline{K}}, \overline{C}_{\overline{K}}, 
\overline{C}'_{\overline{K}} \right)
\left( \begin{array}{ccc} m_K & 0 & \langle \overline{C} \rangle
\langle \overline{P} \rangle/\sqrt{2} M_P \\ 0 & 0 & \alpha_K 
a \langle \overline{P} \rangle/M_P \\ \langle C \rangle \langle P \rangle 
/\sqrt{2} M_P & \alpha_K a \langle P \rangle/M_P & 0
\end{array} \right) \left( \begin{array}{c}
A_K \\ C_K \\ C'_K \end{array} \right).
\end{equation}

\noindent
Here $\alpha_K \equiv \frac{3}{2} ((B-L)_K -1)$, and takes the values
$-1$, $-2$, and $0$ respectively for $K = (3,2,\frac{1}{6})$, 
$(\overline{3}, 1, - \frac{1}{3})$, and $(1,1, +1)$. The entry
$m_K$ vanishes for the color-triplet values of $K$, since $W_A$
has goldstone modes in those directions, but is non-zero (and in fact
equal to $a^2/2M$) in the $(1,1, +1)$ direction. Thus for each $K$
the $3 \times 3$ mass matrix has one vanishing eigenvalue, corresponding
to a goldstone mode that gets eaten by the Higgs mechanism, and two
non-vanishing GUT-scale eigenvalues. One sees also that the massless
mode for $K = (1,1, +1)$ is purely in the $C$ direction, as it should be 
since only the spinor VEVs
break that generator, while for the mass matrices of the color-triplet
representations the massless mode is a linear combination of the adjoint
and spinor as it should be. 

As for the representations $(1,2, - \frac{1}{2})$ and $(\overline{3},1,
\frac{1}{3})$ that are contained in the $\overline{{\bf 5}}$ of 
$SU(5)$, and their conjugates, they are contained only in the
spinors and obtain mass only from the VEVs of $A$, $Z$, and $\overline{Z}$. 
It is easy to see that
the Weak doublets get mass of $3 a \langle P \; {\rm or} \; \overline{P}
\rangle/M_P$, while the color triplets get
mass of $2 a \langle P \; {\rm or} \; \overline{P} \rangle/M_P$. 

In addition to these, the adjoint contains the $(8, 1, 0)$ and 
$(1,3, 0)$, which get mass of $2 a^2/M$ and $a^2/M$ respectively, and the 
$(3,2, - \frac{5}{6}) + {\rm H.c.}$, which get eaten. There are also
several singlets of $G_{SM}$ which get superlarge mass. 
We have thus seen that
no goldstone or pseudo-goldstones are left after symmetry breaking.

From the explicit spectrum given above one can compute the corrections
to the low-energy gauge couplings due to the superheavy states. 
Since $\sin^2 \theta_W$ and $\alpha$ are better known, it is now usual
to use them as inputs for a prediction of $\alpha_s(M_Z)$. The minimal
$SU(5)$ SUSY-GUT predicts$^1$ $\alpha_s(M_Z) = 0.127 \pm 0.005
\pm 0.002$, where the first error is the uncertainty in the low-energy
sparticle spectrum, and the second is the uncertainty in the masses
of the top quark and Higgs bosons. A global fit$^{14}$ to $\alpha_s$
from measurements at all energies gives $\alpha_s(M_Z) = 0.117 \pm
0.005$, while a global fit to all electroweak data$^{15}$ gives
$\alpha_s(M_Z) = 0.127 \pm 0.005 \pm 0.002 \pm 0.001$.

Using the notation of Ref. 8, we define $\epsilon_3 \equiv
(\alpha_3 (M_G) - \tilde{\alpha}_G)/\tilde{\alpha}_G$, where
$\tilde{\alpha}_G \equiv \alpha_1(M_G) = \alpha_2(M_G)$. ($M_G$
is here defined to be the scale at which $\alpha_1$ and $\alpha_2$
are equal.) To obtain $\alpha_s(M_Z) \simeq 0.12$ requires, in
general, that $\epsilon_3 \sim -(2 \; {\rm to} \; 3) \%$. In the 
minimal $SO(10)$ scheme presented here one finds that
$\epsilon_3 \cong \frac{3}{5 \pi} \tilde{\alpha}_G \ln \left[
\frac{32}{9 \sqrt{2}} \frac{\tilde{M}_T}{M_G} \right]$,
where $\tilde{M}_T = a^2/2\langle S \rangle$ is the effective
color-triplet Higgsino mass that comes into the Higgsino-mediated
proton-decay amplitude. $\epsilon_3$ thus comes out to be
$+0.03$ if $\tilde{M}_T \simeq 10 M_G$, and $+0.06$ if 
$\tilde{M}_T \simeq 10^3 M_G$ (as is typically necessary if 
$\tan \beta$ is large). This is somewhat large, but not necessarily
unacceptable, as there can be contributions from other sectors
of the theory, such as the quark and lepton sector, that could have the
opposite sign. 

The stability of the gauge hierarchy requires that certain operators
allowed by $SO(10)$ not be present in $W$ to sufficiently high order in
$M_P$. These include $T_1^2$, $T_1 T_2$, $\overline{C} A C$, $\overline{C} 
C$, or these factors times $SO(10)$-singlet products of fields 
with non-vanishing VEVs. The first of these types of term can
be prevented by a symmetry under which $T_1 \longrightarrow e^{i \alpha}
T_1$, $T_2 \longrightarrow e^{-i \alpha} T_2$, $A \longrightarrow A$, and $S 
\longrightarrow e^{2 i \alpha} S$. Holomorphy forbids the operator
$T_1^2 S^{\dagger}$, of course, but it is necessary also that there be
no chiral product of superfields with the same quantum numbers as 
$S^{\dagger}$ and having VEV which is greater than order $M_G^5/M_P^4
\sim M_W$. For example, the VEV of $S$ may not arise from terms like
$Y_S(S \tilde{S} - m^2)$, as then $T_1^2 \tilde{S}$ would be
allowed by symmetry. This is not a problem, however, as $S$ may acquire
its VEV radiatively, and there are also ways that it may acquire 
a tree-level VEV without giving rise to dangerous operators.$^7$

The term $T_1 A T_2$ is an essential ingredient of the DW mechanism,
but the term $T_1 T_2 {\rm tr} A^2$ must be forbidden. This can be
done by a $Z_2$ under which $T_1$, $A$, $P$, and $\overline{P}$ are
odd and all other fields even.

In fact, all of the destabilizing operators can be forbidden by a 
$U(1) \times Z_2$ symmetry (or a suitable discrete subgroup of it)
under which the fields have the following charges: $A(0,-)$,
$P_A(0,+)$, $X(x,+)$, $C(c-\frac{x}{2}, +)$, $\overline{C}(-c-
\frac{x}{2}, +)$, $P_C(-\frac{x}{2}, +)$, $P(p,-)$, $\overline{P}
(\overline{p}, -)$, $Z(p,+)$, $\overline{Z}(\overline{p},+)$, 
$C'(c-\overline{p} + \frac{x}{2}, +)$, $\overline{C}'(-c -p + \frac{x}
{2}, +)$, $T_1(-t, -)$, $T_2(t,+)$, $S(2t,+)$. (The full symmetry
of the Higgs superpotential presented above is clearly $U(1)^5 \times
Z_2$.) One can now see why the factors $P$ and $\overline{P}$ must
appear in Eq. (3). Otherwise, $Z$ and $A$ would have the same quantum numbers
and $T_1 Z T_2$ would be allowed.

Destabilizing operators can arise if new fields are introduced to 
allow some of the singlets to acquire VEVs at tree level. For example,
if $\langle P \rangle$ arises from a term $Y( P \tilde{P} - m^2)$, then 
the term $T_1 T_2 Z \tilde{P}/M_P$ is allowed
(since $\tilde{P}$ has the quantum numbers $(-p,-)$). However, as
$\langle Z \rangle \sim M_G^2/M_P$, this gives effectively $(M_G^3/
M_P^2) T_1 T_2$, which in turn gives a ``see-saw" contribution
to the $\mu$ parameter of order $(M_G^3/M_P^2)^2/\langle S \rangle
\sim M_G^5/M_P^4$. This is not only acceptable, but even desirable
as a solution to the $\mu$-problem. (Similarly, $\overline{C}
A C Z \tilde{P}/M_P^2$ would be allowed. It is easy to see that then
$\langle A \rangle = {\rm diag} (b,b,a,a,a) \otimes i \tau_2$, where
$b = O(M_G^3/M_P^2)$. Again, this gives a ``see-saw" contribution to
$\mu$ of order $M_G^5/M_P^4$.)

The VEV of $P_C$ cannot arise, however, from the analogous term
$Y_C(P_C \tilde{P_C} - M_C^2)$, as that would imply that $\overline{C}
C \tilde{P_C}^2/M_P$ was allowed. But a term like
$Y_C( P_C^3 \tilde{P_C} - M_C^4)/M_P^2$ would be safe. $((\overline{C}
C)^3 \tilde{P_C}^2)/M_P^5$ would then be allowed, but is harmless.
 
In conclusion, we have demonstrated a simple supersymmetric SO(10) model
which both breaks SO(10) to the Standard Model and solves the doublet/triplet
splitting problem.  In one version of the model a $\mu$ term is generated
naturally.  There is yet one caveat of our approach. In the simplest
version there are several flat
directions at tree level in which some of the required VEVs must be generated
radiatively once SUSY breaking and perhaps supergravity is included. 

\vspace{1cm}

\section*{References}

\begin{enumerate}

\item S. Dimopoulos, S. Raby, and F. Wilczek, {\it Phys. Rev.}
{\bf D24}, 1681 (1981);
U. Amaldi, W. deBoer, and H. Furstenau, {\it Phys. Lett.}
{\bf 260B}, 447 (1991); P. Langacker and M.-X. Luo, {\it Phys.
Rev.} {\bf D44}, 817 (1991); J. Ellis, S. Kelley, and D.V.
Nanopoulos, {\it Phys. Lett.} {\bf 260B}, 131 (1991);
P.Langacker and N. Polonsky, {\it Phys. Rev.} {\bf D47},
4028 (1993).
\item J.A. Harvey and M.S. Turner, {\it Phys. Rev.}
{\bf D42}, 3344 (1990); G.Lazarides and Q. Shafi, {\it Phys. Lett.} 
{\bf B258}, 305 (1991);
J.A. Harvey and E.W. Kolb, {\it Phys. Rev.} {\bf D24}, 2090 (1981).
\item S.M. Barr, {\it Phys. Rev.} {\bf D24}, 1895 (1981);
{\it Phys. Rev. Lett.} {\bf 64}, 353 (1990); K.S. Babu
and S.M. Barr, {\it Phys. Rev. Lett.} {\bf 75}, 2088 (1995);
G. Anderson, S. Raby, S. Dimopoulos, L.J. Hall, and G.D. Starkman,
{\it Phys. Rev.}
{\bf D49}, 3660 (1994); L.J. Hall and S. Raby, {\it Phys. Rev.}
{\bf D51}, 6524 (1995); V. Lucas and S. Raby, {\it Phys. Rev.}
{\bf D54}, 2261 (1996); {\it ibid.} {\bf D55}, 6986 (197).
\item E. Gildener and S. Weinberg, {\it Phys. Rev.} {\bf D13},
3333 (1976); E. Gildener, {\it Phys. Rev.} {\bf D14}, 1667 (1976). 
\item L. Maiani, in {\it Comptes Rendus de l'Ecole d'Et\`{e} de
Physiques des Particules}, Gif-sur-Yvette, 1979, IN2P3, Paris, 1980,
p.3; S. Dimopoulos and H. Georgi, {\it Nucl. Phys} {\bf B150}, 193
(1981); M. Sakai, {\it Z. Phys} {\bf C11}, 153 (1981); E. Witten,
{\it Nucl. Phys.} {\bf B188}, 573 (1981).
\item S. Dimopoulos and F. Wilczek, Preprint NSF-ITP-82-07 (1982).
\item K.S. Babu and S.M. Barr, {\it Phys. Rev.} {\bf D48}, 5354
(1993); {\it ibid.} {\bf D50}, 3529 (1994).
\item V. Lucas and S. Raby, ref. 3.
\item K.R. Dienes, {\it Nucl. Phys.} {\bf B488}, 141 (1997).
\item K.S. Babu and S.M. Barr, {\it Phys. Rev.} {\bf D51}, 2463 (1995).
\item J. Hisano, H. Murayama, and T. Yanagida, {\it Phys. Rev.}
{\bf D49}, 4966 (1994).
\item V. Lucas and S. Raby, ref. 3; Z. Berezhiani and Z. Tavartkiladze,
hep-ph/9612232.
\item S. Urano and R. Arnowitt, hep-ph/9611389.
\item B.R. Webber, in {\it Proceedings 27th International Conference
on High Energy Physics}, Glasgow, Scotland, 1994, edited by 
P.J. Bussey and I.G. Knowles (IOP, London, 1995).
\item J. Erler and P. Langacker, {\it Phys. Rev.} {\bf D52}, 441 (1995).
\end{enumerate}

\end{document}